# Biometric Cards for Indian Population: Role of Mathematical Models in Assisting and Planning

Arni S. R. Srinivasa Rao
Bayesian Interdisciplinary Research Unit
Indian Statistical Institute
203 B.T. Road, Kolkata 700108
Email: arni@isical.ac.in

**Abstract:** Mathematical models could be helpful in assisting the Indian Government's new initiative of issuing biometric cards to its citizens. In this note, we look into the role of mathematical models in estimating the missing, non-enumerated population numbers, estimating annual numbers of cards required by age, gender and regions in India. The linkage between National Population Register and biometric cards is also highlighted. See technical Appendices. There are other scientific issues, namely, electronic, data storage management, identity verification etc, which we do not address in this paper.

**Acknowledgement:** This note was highly benefited by the comments from Professors N.V. Joshi, Indian Institute of Science, Bangalore and Philip K Maini, University of Oxford. I also acknowledge valuable suggestions from the Chief Editor (Asian Population Studies, Routledge, Taylor & Francis Group) which helped to revise the paper.

## Introduction:

The Indian government has launched a mammoth project of UIDAI (Unique Identification Authority of India) cards (henceforth also referred to as biometric cards) for its citizens **[1, 2, 3]**. The rationale of introducing these cards is for effective implementation of various welfare programs in the country. These welfare programs could range from agricultural, health, financial, family welfare, small scale industries, employment, self-employment schemes to any other non-program related ones. Media reports suggest that it is indeed a very useful project for a country which has an estimated population of 1.2 billion **[4, 5]**. Statistically speaking, the proposed project could in principle provide a better data source on the population in general when compared to the existing PAN cards issued by income tax departments and voter identity cards issued by the election commission, which are issued to individuals above the age of eighteen or individuals who satisfy certain eligibility criteria. The Indian Census is in 2011 and procedures involved for the smooth running of this (for example, questionnaire designs, methods on data collection etc) were launched in 2008 **[6]**. As far as the demographic data are concerned the Census of India offers the largest dataset on the Indian population in terms of coverage and size. The net omission rates in various Indian zones, as per the Census 2001, range between 20 and 49 per 1000 individuals **[7]**.

## Model and Analysis:

Biometric cards project links its operations with National Population Register (NPR) of the Census 2011 operations. During NPR process, in addition to collecting demographic, location information, such as, name, age, gender, address, etc, biometric information, such as digital photograph, finger prints of both the hands, iris will be collected. Later, this data will be used to remove duplicated entries before generating unique identity numbers for each individual in every household. Above biometric information is collected for persons aged 15 and above and for persons below age 15 unique identity numbers will be associated with parent or guardian as their biometric information is not collected. All the data unique identity numbers will be also maintained at office of the Registrar General of India.

The strategies outlined in the draft by the UIDAI **[1]**, does take care necessary points for a successful implementation of biometric cards. These include, advantages of having a UID by citizens, data standards, costs involved in implementation, type of the demographic data to be collected, enrolment strategy in rural and urban areas, security of the cards and data, linking of biometric card project with existing government and nongovernment agencies, projected revenues generated by the project etc.

Other than the other advantages mentioned elsewhere in this perspectives, biometric cards can be utilized by citizens during natural calamities like floods, train and road accidents, flight accidents, etc for producing and or matching biometric identities. The other intended users of these cards are various government, semi-government agencies, banks etc. The biometric cards project can utilize the census operations to maximize the accuracy in distribution of cards, reduction of duplicated cards, etc, Eventually when the biometric card data become complete they can be used to fill the gaps in population counts. Undoubtedly these biometric cards will help us in conducting policy based research in areas including health, economy, military recruitment, etc,

There could be potential disadvantages that arise due to implementation of cards for public service. Governments need to assess possible disadvantages of computerization of key citizen database.

There is a possibility that the data will be utilized by the Indian and foreign commercial business programs.

However, at the same time; there are key scientific issues which need to be carefully addressed by the UIDAI. These can be broadly classified into four categories: First, the clarity on utility of these cards over existing cards or synchronization of existing card data with the proposed cards; second, how much information is to be collected with a certain degree of accuracy, design of electronic component of the biometric card; third, costing of the proposed project or costs involved in successful implementation; fourth, scientific design for reaching all citizens to collect their data and issuing of cards. All these issues need to be addressed using scientific methods so as the mapping of citizens living in the country and outside is complete. The fourth category involves tackling ground-level difficulties, namely, accurate estimates of counts of the populations by region, state, block, ward etc, recruitment of investigators for data collection, data entry operators, language issues for sorting out the communication gap between top level scientists to the ground level scientific workers involved in the project. If any of the bottom level work is carried out in an improper way then it could lead to issuing of cards to ineligible persons and missing out some citizens.

Mathematical models and micro-simulation models have a potential role in assisting the entire program in general and in particular the fourth category discussed above. The importance of models can be classified in three ways, a) for simulating the citizens' data by regions, states, blocks and wards with a combination of required demographic and social structure (this will serve as a background for the study), b) models can use the data collected by UIDAI and then help in estimating missing citizens in the program, c) once the biometric cards are issued to all citizens (with 100 percent enumeration), then models can incorporate the dynamic nature of the population to estimate annual new requirements of the cards and annual number of cards that are to be destroyed (which will arise due to deaths, emigration, other causes etc) in advance. The mathematical techniques involved could range over a variety of topics, such as, dynamical systems, discrete mathematics, deterministic algorithms, capture-recapture techniques, etc, which require high performance computing machines. In this brief "perspective" we will not be dealing with administrative issues such as the mechanism required for issuing new UID cards to eligible persons each year.

The government also needs to address further important questions, are these new biometric cards going to be helpful to construct population registration (by identifying illegal migrants and denying cards for them etc) in the country or alternatively, will the accurate population registration (or population counts of bona-fide citizens) help in issuing biometric cards? Models can also assist in deriving an answer for both types of questions depending on what the government intends to achieve.

Indian population (provisional) for the year 2011 is 1,210 million (623 million males and 586 million females) **[8].** Each year the number of male and female cards required after 2011 are shown in Figure 1. Projected decadal population growth rate from the next Indian census i.e. 2011 to 2021 is 12.3 percent. Models can help to estimate the babies that will be born in each year by gender, by regions, by states in India in the near future with some degree of accuracy, which can be useful for planning and distribution of cards. (See Appendix I & II for a prototype model to estimate births by region and gender). There is a separate report by Census of India on population projections up to the year 2026 by a technical group on population projections [9]. These projected populations are valid, however in terms of biometric cards estimate, these could be an over estimate if we use these numbers as number of requirements of the biometric cards. This is due to the fact that cards will be issued only for eligible people which could be less than or equal to the actual people living in the country. In Appendix III we provide the idea of approximating annual number of new cards required. According to the 2001 census, there are about 0.3 percent of the people who have not reported their age **[7].** Unknown age or unknown date of birth of individuals is a concern and obtaining correct age of all the citizens at the time of issuing the cards is a challenging task. Conversely, biometric cards data can be used indirectly for demographic analysis, for example, estimating population movement or specifically migration information (if addresses are regularly updated), estimating annual age specific deaths (in case of availability of data on number of cards returned due to death of persons or returning card numbers in case deaths below age 15). There needs further clarity on method to be followed for returning of cards of died persons. Since newly born and children up to age 14 are linked to parents or guardian cards, there requires procedures to remove numbers of aged persons from age 14 to age 15 and issuing of new cards to age 15 persons to avoid arising of duplicate cards. Models can be constructed to ascertain the degree of age-misreporting, but they often fail to impute the exact age of the citizens. The distribution of population by age is available and this information across the country by region can also be obtainable. For example, there were about 303 million males and 282 million females in the age group of 15-59 in the census 2001. These values for the age '60 and above' were 38 million and 39 million for males and females respectively. Interestingly, in the census 2001, there were about 1.5 million males and 1.2 million females whose age was not stated. Mathematical models can be developed for projecting the population age-sex distribution, which will assist UIDAI program. Issuing biometric cards to population who do not have census defined houses needs further careful approach. There are a sizable number of citizens in urban and rural areas who live in one or more of the places, namely, pavements, railway stations, under fly over bridges, etc. For example, according to 2001 Census, there were about 1.2 million houseless individuals in rural and 0.8 million houseless individuals in urban areas. Modeling these numbers for assisting cards program requires new techniques and identification of such population is a challenging task for the UIDAI team.

Information on migration (temporary as well as permanent) within states and between states is very important, if government requires reaching the citizens at the time of rescue operations, implementation of pubic distribution systems (PDS), flood affected areas, relocation due to industrialization etc. For example, using of not updated population movement data could lead to inaccurate assessment of the actual number of citizens living in a geographical area. Hence, there could be possibility of disproportionate funds are distributed to service operation. Rural and urban planning, like, setting up of economic zones in specific cities or villages an annual updating of population movement is important in handling dynamic nature of the Indian population activities.

## Summary:
Biometric cards project, once implemented, is expected to enroll 200 million citizens in 2-3 years and 600 million in 4-5 years **[1]**. Enrolled data can be used by mathematical models for estimating annual new born babies (who are to be given UID) by using information such as age, gender, marriage, reproductive statistics etc (Note that UIDAI demographic data do not propose to collect marriage status and it might be useful to collect). We agree that the quality of the any survey or census data depends, among other things, on seriousness of the respondents while handling the questions and enumerators while taking their role. Similarly the biometric data is expected to get finer over the years of implementation of the project and after necessary correction measures like, de-duplication etc, Finer data can be successfully used in deterministic models, for example, to computation of dynamics of population, migration dynamics within and between Indian states, the population relocation process due to economic activity, industrialization, etc, to support other developmental and services by the governments. It is also important to take experience from countries like UK and USA who have abandoned such projects in the past **[10].**

# Appendix I

**Estimating the Annual Requirements of new UIDAI cards in terms of new Births by Gender and Region**

Let $P_t(x)$ represent projected population of age *x* at time *t*, *s(x, x+1)* is a survival function, which describes the chance of survival for an individual of age *x* to age *x+1* and *F(x)* is age specific rate of reproduction of new babies at age *x*.

Then

$$\sum_{x=15}^{49} S(x, x+1) P_t(x) F(x) K$$

Gives number of newly born babies that will be delivered by mothers aged between 15 and 49. Here, *K* is the proportion of eligible female out of total female in the reproductive age group. Since we assume every newly born live birth will be given a biometric card, we do not consider modelling the deaths of children.

In a closed population, if $P_0(x; n, r)$ is female population at time *t=0* in the age group $(x, x+n)$ in a region *r*, where $x = 0, 1, 2, \ldots, \omega$. $r = 0, 1, 2, \ldots$ . Here, $\omega$ is last age of life. Then the number of citizens who will be alive at the end of *z* years in the region *r*, i.e. $p_z(x+z; n, r)$ is

$$P_z(x+z; n, r) = \int_0^n p(x+t; r) s(x+z, t) dt$$

Where $p(x; r)$ is no. of citizens of age *x* to *x+dx* in the region *r*, and $s(x+z, t)$ is chance of survival of a citizen of age *x+z* survives to age *x+ t + z*. Models using survival functions are used for population projections **[11]**. In case the cards are to be issued to children who survive to age one, (i.e. excluding the infant deaths, which is 60 per 1000 live births), then above equations can be incorporated by survival factor for a live birth to complete one year.

# Appendix II

**Model for Estimating the Population Survived from Currently Living Population: A difference equation approach.**

Using a simple difference equation, we can obtain the future survivors out of current living population. Suppose $P_r^{(t_1)}(x, x+1)$ is the population available at census time point $t_1$ and for region *r* for the ages $x = 0, 1, 2, \ldots, \omega$. We can obtain the population who are surviving out of $P_r^{(t_1)}(x, x+1)$ at the census time point $t_2$ by using the survival function, $S_r^{(t_2-t_1)}(x, x+1)$, representing the chance of survival of population of age $x$ in a region $r$ at census time $t_1$ to live up to age $x+(t_2-t_1)$ at census time point $t_2$. Such functions for all the ages can be computed if a life table is available for the region $r$. Thus the population that survives throughout the time period $t_2-t_1$ and reaches the census time point $t_2$ at age $x+(t_2-t_1)$, i.e. $P_r^{(t_2)}(x+(t_2-t_1), x+1+(t_2-t_1))$ can be expressed as,

$$P_r^{(t_2)}(x+(t_2-t_1), x+1+(t_2-t_1)) = P_r^{(t_1)}(x, x+1) S_r^{(t_2-t_1)}(x, x+1),$$ for the values of $x=0,1,2,\ldots,\omega-1$. For further information on difference equation models in biology see **[11, 12]**.

# Appendix III

**Macro level and Micro level Model for Change in annual UID requirements.**

Suppose total UID cards are provided to the eligible persons in all the states and union territories in India. Accurately predicting the time required to issuing cards to the existing persons is a challenging issue, because of several administrative factors involved. These times could vary across the states and the union territories. For example, theoretical modeling of the time required to vaccinate against epidemics indicates, that there could be variations in times in the sub-populations in the country **[13]**. Moreover, we have not yet known the strategies that are implemented in finally distributing the biometric cards to the citizens. Hence, we discuss here macro level and micro models that describes the process of arrival of a new card eligible member annually. Let $U$ be the total cards issued, then net change in the cards in the next year are

$$\frac{dU}{dt} = \int b(i)S(i)di - \int d(i)S(i)di + \int m(i)S(i)di - \int e(i)S(i)di$$

Where *b(i)* is the rate of newly born people, *d(i)* is the rate of death of card holders, people enter into the state *i* (from out side the country) at a rate *m(i)* and leave the country at a rate $e(i)$. This is a macro level model, where *b, d, m, e* are state or a union territory specific. Since new cards are needed for those who are either born or entered into the country with having eligibility of having a card, we can consider rate of change in the new cards as

$$\frac{dU}{dt} = \int \left[b(i) + m(i)\right] S(i) di$$

Suppose we wish to look into a micro level scenario of a state. Let $S_{ni}$, $S_{mi}$, $S_{di}$, be the $n^{th}$, $m^{th}$, $d^{th}$, $e^{th}$ person who are born, entered, died, in the state *i* respectively and $S_{ei}$ be the $e^{th}$ person who left state *i* and entered another state *j* in a year. Let $S_{fi}$ be the $f^{th}$ person who left the country and $S_{gi}$ be the $g^{th}$ person who entered into the state *l* from outside the country. Then the net number of change in the cards in the next year (given there are *U(S_i)* cards issued in the previous year) are

$$\frac{dU(S_i)}{dt} = \sum_i S_{ni} - \sum_i S_{di} + \sum_i S_{mi} - \sum_i S_{ei} + \sum_i S_{gi} - \sum_i S_{fi}$$

Here $\sum_i S_{ei}$ are number of total people who have moved outside the state *i*. This number will be distributed into one or more number of newly added cards in respective states i.e. it contributes to at least one of the $S_{mj}$ for $i \neq j$. Similarly $\sum_i S_{mi}$ is contributed by at least one of the *j* states for $i \neq j$.

As above, equation for the change in the number of cards those are to be issued annually is

$$\frac{dU(S_i)}{dt} = \sum_i S_{ni} + \sum_i S_{mi} + \sum_i S_{gi}$$

Such a micro level modeling analysis will help to track the village or block level change in the populations in the country. Once the databases of the cards are released, several other causes of migration by economic activity etc can be obtained at a grass root level. Suppose β is vector of parameters describing the annual requirement of the biometric cards, then the information available on births, deaths and migration (say, *D*) can be utilized for updating the prior distribution of β, (say, *p*(β), which is suitable chosen), via likelihood function (say, P(*D*/β)). Using the Bayesian framework, one can estimate posterior distribution, *f*(β/*D*) as follows:

$$f(\beta/D) = \frac{P(D/\beta)p(\beta)}{\int P(D/\beta)p(\beta)d\beta}$$

However, before updating the prior distribution, problem lies in experienced selection of a suitable prior distribution, because *p*(β) is conditional distribution of β before the data *D* is collected. Depending upon the accuracy of the estimate, mean of the posterior distribution can be adapted as a point estimate of the random variable of interest. The integral in the above equation can be solved suing Markov chain monte carlo computational methods (see details of such computational techniques for example in [14].

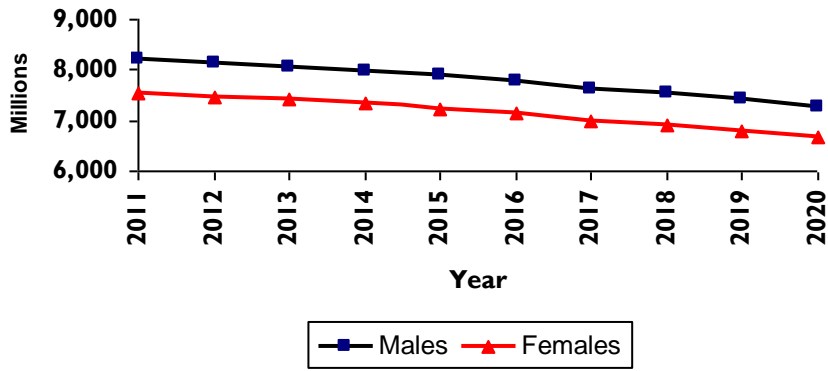

**Figure 1. Predicted Annual Number of Cards Newly Required**